# Synchronization Stability Analysis and Enhancement of Grid-Tied Multi-Converter Systems

Xiuqiang He, Hua Geng

Beijing National Research Center for Information Science and Technology, Department of Automation
Tsinghua University, Beijing, China
he-xq16@mails.tsinghua.edu.cn

*Abstract*—Synchronization instability of grid-tied converters during grid faults is a serious concern. For phase-locked loop (PLL) synchronized converters, synchronization stability during grid faults requires that PLL can resynchronize with the power grid. It is of great significance for meeting grid code specifications such as reactive current support. However, the synchronization in the context of high-impedance weak grid is not always readily achieved. On one hand, the terminal voltage of converters is changeable due to being susceptible to the output current. One the other hand, there is interaction between multiple converters interconnected to the grid. Therefore, loss of synchronism (LOS) of converters during grid faults likely occurs. Previous research has investigated the LOS issue of single-converter infinite-bus (SCIB) system, but mostly overlooked the interaction between multiple converters. Previous efforts to address the issue of LOS are also in doubt for multi-converter systems. This paper develops a multi-converter infinite-bus (MCIB) system model and analyzes the root cause of LOS with considering multi-converter interaction during grid faults. Moreover, a feedforward compensated PLL (FFC-PLL) method applicable for multi-converter systems is developed to address the LOS issue. Simulations results verify the effectiveness of the method.

## I. INTRODUCTION

Voltage source converters (VSCs) have been widely used as interface of renewable energy generation systems to connect with the power grid. As the power grid becomes increasingly weak, instability issues of grid-tied VSCs gradually emerge, generally including small-signal instability and transient instability. Small-signal stability analysis of VSCs have been extensively investigated [1], [2]. However, transient stability research of VSCs still faces great challenges [3], [4].

Transient stability characteristics of conventional power systems are shaped by physical motion of rotor of synchronous generators [5]. Hence, conventional transient stability attributes to rotor angle stability [6]. Different from synchronous generators, transient stability characteristics of VSCs are primarily determined by control strategies (including protective measures) that can be customized to suit specific needs [7]. Especially for the most commonly-used VSCs that are synchronized with the grid through phase-locked loop (PLL), their transient stability characteristics are largely shaped by the character of PLL [8]–[14]. Transient stability of this type of VSCs is also termed as synchronization stability [8]. Specifically, synchronization stability concerns about whether VSCs can successfully synchronize with the grid via PLL when subjected to disturbances such as grid faults. The synchronization is crucial for VSCs to prevent tripping off during grid faults, thereby remaining connected with the grid to provide ancillary services as per grid codes [15], such as perform reactive power support. However, it is not always easy for VSCs connected to a high-impedance weak grid to achieve this purpose [8]–[14], [16]–[19].

For a VSC connected to a high-impedance weak grid, its terminal voltage is susceptible to output current. Besides, there are often plentiful VSC-interfaced generation units in a single or multiple generation plants and they are interconnected by collector line to grid-connection point (GCP). There is usually interaction between multiple VSC units since the terminal voltage of one unit is sensitive to the output current of the others. The multi-converter interaction makes the terminal voltage dynamics much complicated. In this case, PLL may not track the angle of the terminal voltage and consequently fail to synchronize with the grid, which is termed as loss of synchronism (LOS) [16], [17]. Tripping events of photovoltaic converters across a wide area have been reported and it has been found that such events were associated with LOS [20], [21].

When subjected to severe grid faults, the terminal voltage of VSCs dips significantly. Provided that there is a (possibly new) steady state, synchronization stability expects that state variable of PLL can successfully transit from the pre-fault steady state to the post-fault one. Previous research efforts [8]–[14], [16]–[19] have made significant achievements in this respect, although they were still not extensive and thorough. At the first step, single-converter infinite-bus (SCIB) system was taken as study object for simplicity. References [17] and [18] made pioneering efforts and developed a stability criterion regarding the existence of steady state equilibrium point after grid faults. A more extensive investigation was made in [19] by considering different operating conditions. Based on findings in [17] and [18], the latest research [8]–[14] in the last few decades have revealed that synchronization stability was not only related to the existence of post-fault equilibrium point but also affected by the location of initial state and the dynamic performance of PLL. To analyze dynamic process of PLL and predict the synchronization stability after being subjected to grid faults, motion equation analysis [8], equal area criterion [9]–[11], and phase portrait method [12]–[14] have been utilized in previous research. The research findings have revealed the fundamental mechanism of synchronization stability of SCIB system. Nonetheless, the impact of multi-converter interaction remains unclear currently, which is the first focus of this paper.

Most grid-tied VSCs in current industrial applications may still not have a simple solution except tripping to avoid LOS (i.e., synchronization instability) [20]. In fact, academia has developed several solutions, basically including four typical methods. The first one is the most straightforward method that is to freeze PLL's regulator once PLL frequency significantly deviates from the fundamental frequency [16]. It must be noted that the method has the problem of static error since the PLL after being frozen loses tracking capability [10]. Still, the method was recommended by NERC [20]. The second method [10] proposes to retain PLL's proportional regulator while freezing/removing integral regulator during grid faults. But, the availability of the method requires that there must be equilibrium point after faults [10]. The third method [8] is to adjust ratio of reactive/active current output based on grid impedance angle. However, it is difficult to fast detect the angle of post-fault grid impedance. The fourth method [17] proposes to regulate active current output according to the deviation of PLL frequency. It is found in this study that it seems to be hard to predict its steady-state performance. The four methods have various disadvantages affecting their performance. More importantly, the applicability of the latter two methods in multi-converter systems were doubtful because they were proposed from the standpoint of SCIB system. Hence, it is necessary to explore more superior solutions applicable for multi-converter systems, which is the second focus of this paper.

This paper develops a multi-converter infinite-bus (MCIB) system model based on the existing model of SCIB system [8]. On one hand, the model of a certain VSC in MCIB system suggests that the output current from itself and the others together produces an offset term in its own PLL input. It adversely affects the stability of itself and responsible for potential LOS events. On the other hand, the output currents of multiple VSCs gather together at intersections of collector network and then flow through common impedance. Thus, multi-converter interaction arises on the common impedance. The multi-converter interaction amplifies the magnitude of the offset term in PLL input and consequently increases the risk of LOS. With the increase of grid impedance, the interaction becomes intense and especially remarkable in high-impedance weak grid. Considering that the offset term is detectable, a feedforward compensated PLL method is developed to eliminate the adverse effect of the offset term so as to address the issue of LOS. The method is applicable for multi-converter systems. Its performance is verified by comparisons with existing methods.

The rest of this paper is organized as follows. Section II develops the model of MCIB system. Section III analyzes the root cause of LOS and the impact of multi-converter interaction on LOS. Section IV develops a feedforward compensated PLL, which is verified by simulations in Section V. Section VI concludes this paper.

## II. SYSTEM MODELLING

Since the model of MCIB system takes that of SCIB system as the basis, the model of the latter will be built firstly, then based on which the model of the former will be developed.

### A. SCIB System Modeling

The circuit and control diagram of SCIB system is illustrated in Fig. 1. Seeing from grid-connection point (GCP) towards the external grid, Thévenin-equivalent grid parameters are ugabc, Rg and Lg. $R_t$ and $L_t$, $R_l$ and $L_l$ are transformer, line impedance parameters, respectively. $u_{abc}$ is terminal voltage and $i_{abc}$ is output current. The reactive current reference $i^*_{q\_fault}$ during grid faults is designated by specifications of grid codes

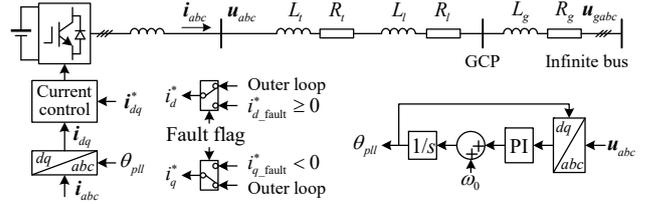

Fig. 1. Circuit and control diagram of SCIB system, which adopts grid-following control mode during severe grid faults to avoid over-current.

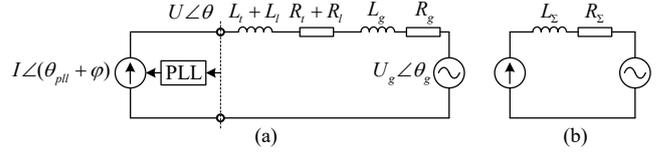

Fig. 2. (a) Circuit of SCIB system where converter is represented by a controlled current source [10]. (b) Equivalent circuit to (a).

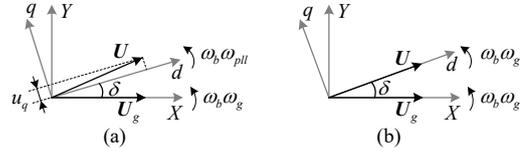

Fig. 3. Relation between $XY$ reference frame and PLL reference frame in (a) dynamic process, and (b) steady state, in both of which $\delta = \theta_{pll} - \theta_g$.

[15]. The active current reference $i^*_{d\_fault}$ has not yet been designated by current grid codes in use and it could be set to zero in view of limit of converter capability [16]. The DC link of converter during grid faults could be modeled as a DC voltage source considering that activated chopper circuit is able to maintain DC-link voltage [8]–[14], [17]–[19].

Grid-tied VSCs typically adopt voltage-oriented current vector control (also termed as grid-following control) during severe grid fault in order to make current properly controlled and inject specific reactive current. Synchronous reference frame based PLL (SRF-PLL) [22] is commonly used for vector orientation in current control. The bandwidth of current control (~200 Hz) is generally much higher than that of PLL (~20 Hz). Therefore, from the VSC terminal point of view and focusing on the time scale of PLL, current control loop together with terminal filter can be simplified to a controlled current source [8]–[14], as shown in Fig. 2. Besides, electromagnetic transient of current is overlooked and thus the circuit can be represented in a form of phasor [10].

The following modeling is based on the per-unit system, in which $\omega_b$ is base value of frequency. The modeling is made in synchronous reference frame for avoiding inconvenient analysis to AC variables. Two reference frames are defined here. The one is infinite-bus ($XY$) reference frame that is rotating with grid frequency $\omega_b\omega_g$. The other is PLL ($dq$) reference frame that is rotating with PLL frequency $\omega_b\omega_{pll}$. As shown in Fig. 3, $d$-axis of PLL reference frame coincides with terminal voltage vector in steady state. The included angle between the two reference frames is represented by $\delta$,

$$d\delta/dt = \omega_b\left(\omega_{pll} - \omega_g\right) \triangleq \omega_b\Delta\omega. \tag{1}$$

The frequency dynamics of SRF-PLL is

$$d(\omega_b \Delta\omega)/dt = d(\omega_b \omega_{pll})/dt = k_p\, du_q/dt + k_i u_q \qquad (2)$$

where $k_p$ and $k_i$ are gains of PLL's regulator and $u_q$ is $q$-axis component of terminal voltage. Regarding (2), it is worth noting that grid frequency dynamics is overlooked considering that it is much slower than PLL frequency dynamics.

Neglecting electromagnetic transients, the voltage equation of the circuit can be written in PLL reference frame as follows:

$$\dot{U} = U_g \angle(-\delta) + Z_\Sigma \dot{I} \qquad (3)$$

where $\dot{U} = u_d + j u_q$, $\dot{I} = I \angle \varphi = i_d + j i_q$, $Z_\Sigma = R_\Sigma + j\omega_{pll} L_\Sigma$.

Equation (3) yields $q$-axis voltage equation:

$$u_q = -U_g \sin\delta + R_\Sigma i_q + \omega_{pll} L_\Sigma i_d \triangleq -U_g \sin\delta + a \qquad (4)$$

which indicates that there is an offset term in $u_q$ and it is caused by the voltage drop across lumped impedance. During normal condition, $a$ is larger than zero because $i_q = 0$ and $i_d > 0$; During fault condition, $a$ probably becomes smaller than zero because $i_q < 0$ and $i_d \geq 0$ [16], as shown in Fig. 1.

Equations (1), (2), and (4) form the nonlinear model of SCIB system, as shown in Fig. 4(a). The model has been widely acknowledged in previous research [8]–[14].

### B. MCIB System Modeling

Different from the single form of output current in (4), there are multiple output currents in a multi-converter system (see Fig. 5), which results in multi-converter interaction. Note that all the output currents flow through grid impedance, and consequently there is interaction between any two VSCs.

For a certain VSC ($VSC_k$) located in any position, Fig. 5 illustrates the effect of output currents from the other ones. There are $N$ VSCs. $n$ VSCs locate in the branch including $VSC_k$. The other $N - n$ ones locate in other branches or plants.

The current phasor of each VSC is oriented by its own PLL reference frame. To highlight this point, the current of $VSC_j$ is denoted by $\dot{I}_j^j$ where the superscript $j$ is to indicate the orientation by its own PLL. Further, $\dot{I}_j^j$ can be transformed to a counterpart oriented by the PLL reference frame of $VSC_k$, using the superscript $k$ to represent the counterpart,

$$\dot{I}_j^k = e^{j(\delta_k - \delta_j)} \dot{I}_j^j \qquad (5)$$

where $\delta_k$ denotes the included angle between PLL reference frame of $VSC_k$ and $XY$ reference frame, similar meaning for $\delta_j$. In Fig. 5, $Z_k$ denotes line impedance and $Z_{kt}$ denotes transformer impedance. Thus, the terminal voltage of $VSC_k$ can be expressed in its own PLL reference frame as:

$$\dot{U}_k^k = U_g \angle(-\delta_k) + Z_g \sum_{j=1}^{N} \dot{I}_j^k + \sum_{i=1}^{k} Z_i \sum_{j=i}^{n} \dot{I}_j^k + Z_{kt} \dot{I}_k^k \qquad (6)$$

After splitting (6) into $d$- and $q$-axis expressions, $q$-axis component $u_{kq}$ of $VSC_k$'s terminal voltage is given by,

$$\begin{aligned} u_{kq} =\ & -U_g \sin\delta_k + R_g \sum_{j=1}^{N} i_{jq}^k + \sum_{i=1}^{k} R_i \sum_{j=i}^{n} i_{jq}^k + R_{kt} i_{kq}^k \\ & + \omega_{kpll}\left( L_g \sum_{j=1}^{N} i_{jd}^k + \sum_{i=1}^{k} L_i \sum_{j=i}^{n} i_{jd}^k + L_{kt} i_{kd}^k \right) \\ \triangleq\ & -U_g \sin\delta_k + a_k \end{aligned} \qquad (7)$$

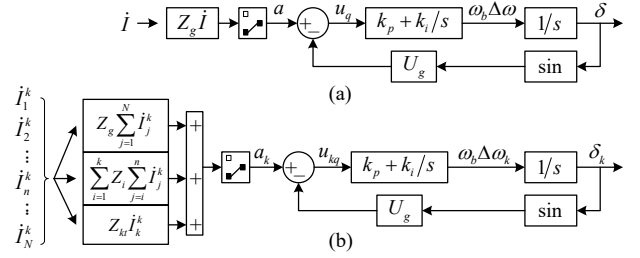

Fig. 4. Nonlinear system models. (a) The model of SCIB system. (b) The model of MCIB system, where $a_k$ quantifies the effect of output currents of a certain VSC oneself and the others.

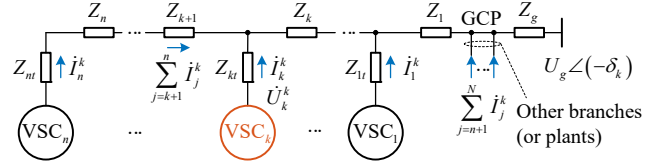

Fig. 5. Circuit expressing the effect of currents from other VSCs on $VSC_k$. Note that the voltage and current phasors in the figure are relative to $VSC_k$'s PLL reference frame.

Equation (7) indicates that there is an offset term $a_k$ in $u_{kq}$ as well, but it is in form of aggregation. The offset term $a_k$ quantifies the combined effect of output currents from $VSC_k$ itself and other VSCs. It is affected by many parameters and its expression is much more complex than the offset term $a$ in (4).

Besides the terminal voltage equation, $VSC_k$'s model also includes its own PLL's dynamics,

$$\begin{aligned} d\delta_k/dt &= \omega_b\left(\omega_{kpll} - \omega_g\right) = \omega_b \Delta\omega_k \\ d(\omega_b \Delta\omega_k)/dt &= k_p\, du_{kq}/dt + k_i u_{kq}. \end{aligned} \qquad (8)$$

The model of $VSC_k$ as a subsystem of MCIB system is shown in Fig. 4(b). Except for the front end, the rest of the model has the same structure with the model of SCIB system.

Here, $n = N = 2$ (i.e., a single-branch two-converter system) is taken as an example to illustrate multi-converter interaction. Equation (6) further yields the following two terminal voltage equations:

$$\begin{aligned} \dot{U}_1^1 &= U_g \angle(-\delta_1) + (Z_g + Z_1)(\dot{I}_1^1 + \dot{I}_2^1) + Z_{1t} \dot{I}_1^1 \\ \dot{U}_2^2 &= U_g \angle(-\delta_2) + (Z_g + Z_1)(\dot{I}_1^2 + \dot{I}_2^2) + (Z_2 + Z_{2t}) \dot{I}_2^2 \end{aligned} \qquad (9)$$

where $\dot{I}_1^1$ and $\dot{I}_2^2$ are output currents of $VSC_1$ and $VSC_2$, respectively; $\dot{I}_2^1 = e^{j(\delta_1 - \delta_2)} \dot{I}_2^2$ and $\dot{I}_1^2 = e^{j(\delta_2 - \delta_1)} \dot{I}_1^1$ are counterparts of $\dot{I}_2^2$ and $\dot{I}_1^1$ after reference frame transformation, respectively.

The diagram of the two-converter system is depicted in Fig. 6(a). The model diagram in Fig. 6(b) suggests that the effect of output current on terminal voltage include two parts: self-effect part and mutual effect part. The self-effect part acts on lumped impedance through which each current flows. The mutual effect part acts on common impedance through which both currents flow. As Fig. 6(b) shows, the mutual effect part represents multi-converter interaction. In contrast, there is only self-effect part in the model of SCIB system.

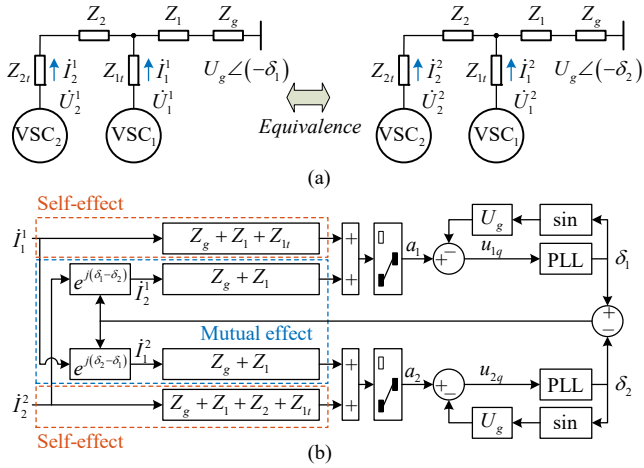

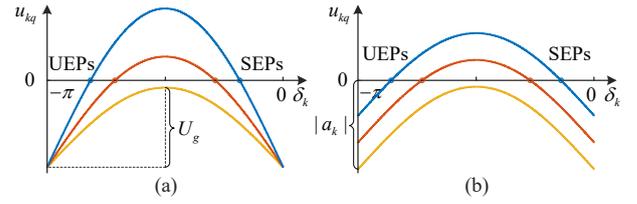

Fig. 6. Single-branch two-converter system and its model. (a) Circuit diagram. (b) Model diagram. Note that the capacity of the VSCs could be rescaled to represent generation plants containing a plenty of VSC units.

## III. Synchronization Stability Analysis

Generally, asymptotic stability of nonlinear systems requires: i) stable equilibrium point (SEP), ii) proper initial state, and iii) good enough dynamic performance. In view of this, stability analysis in this section will be conducted from these three aspects of stability requirements.

As stated in Section II, all VSCs adopt grid-following control mode during grid faults, and therefore their models resemble each other. The right part in Fig. 6 displays that both VSCs have the same model structure, which is with $a_k$ as input variable. It is reasonably believed that there is no essential difference in their fundamental stability mechanism, which makes sense because of adopting the same control mode. In view of this, from a single VSC point of view, at the first step, $a_k$ is regarded as input variable of model structure to clarify the fundamental stability mechanism and root cause of LOS. From multi-converter interaction point of view at the second step, the effect of $a_k$ will be analyzed to reveal the impact of multi-converter interaction on the synchronization stability.

### A. Stability Analysis of $VSC_k$

*1) Stable Equilibrium Point Analysis*: According to (7) and (8), the equilibrium point of subsystem $VSC_k$ is decided by,

$$u_{kq} = 0 \tag{10}$$

which is zero-crossing point of the curve $a_k - U_g\sin\delta_k$, i.e.,

$$a_k = U_g \sin \delta_k. \tag{11}$$

Fig 7 shows several examples of equilibrium point. It is demonstrated in [8] that right zero-crossing points are stable equilibrium points (SEPs) while left zero-crossing points are unstable equilibrium points (UEPs). From Fig. 7, it can also be found that if a severe grid fault or a significant grid impedance causes $|a_k| > U_g$, there will be no longer post-fault equilibrium point. Since $u_{kq} = 0$ is a fundamental condition of voltage orientation in grid-following control mode, it is impossible under such circumstances to achieve synchronization, whether using PLLs [22] or frequency-locked loops (FLLs) [23].

*2) Initial State Analysis*: In the first several cycles after grid

Fig. 7. Equilibrium point of subsystem $VSC_k$. There may be no equilibrium point (a) if the post-fault grid voltage is too small, or (b) the absolute value of offset term is too large.

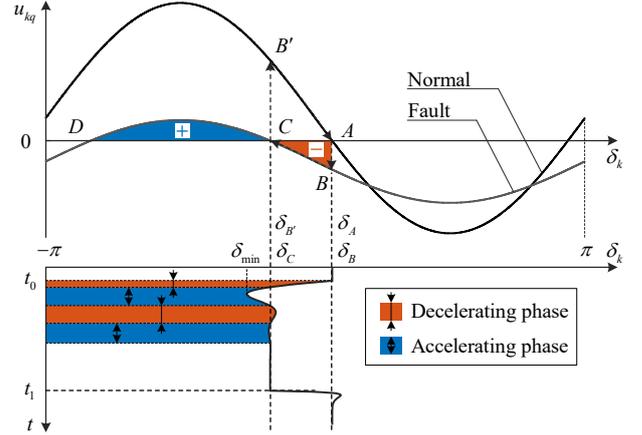

Fig. 8. Dynamic synchronization process after a grid voltage dip and recovery. The offset term in $u_{kq}$ makes the curve of $u_{kq}$ asymmetric with respect to the horizontal axis.

faults, protective measures are activated for protecting VSC devices from damage. Thereafter, VSCs regain control function and enters fault ride-through stage. The beginning time of this stage is called initial time. The initial state of subsystem $VSC_k$ at initial time is denoted as $\delta_{k0}$, which is determined by both initial angle of PLL and initial phase of post-fault grid voltage. It has been remarked in [8] that both are stochastic. On one hand, the initial state of PLL is uncertain because of detection and protection delay. On the other hand, since there is a phase jump at the moment of grid faults, the initial grid phase is unpredictable and it is affected by stochastic factors such as fault location and grounding resistance. Hence, the initial state $\delta_{k0}$ is stochastic, which makes it difficult in practice to assess whether $VSC_k$ can achieve synchronization during grid faults.

*3) Dynamic Performance Analysis*: Focusing on SCIB system, Fig. 8 illustrates dynamic synchronization process after a grid voltage dip and recovery. Points $A$ and $C$ are SEPs whereas $D$ is a UEP. The dip of grid voltage happens at $t_0$ and operating point steps from $A$ to $B$ at this moment. Since $u_{kq}$ becomes smaller than zero, the system enters a decelerating phase and the operating point moves towards $C$. PLL undergoes a decelerating process and $\delta_k$ decreases as well, as shown in Fig. 8.

When the operating point reaches $C$, $\Delta\omega_k$ is smaller than zero because of the preceding decelerating. Consequently, $\delta_k$ will continue decreasing, which leads to angle overshoot. Since $u_{kq}$ becomes larger than zero, the system enters an accelerating phase and $\Delta\omega_k$ begins to increase. When $\Delta\omega_k$ increases to zero, $\delta_k$ reaches the minimum $\delta_{\min}$. Afterwards, an attenuated convergence process is observed. Due to damping effect of PLL's proportional regulator [10], the operating point ultimately converges to *C*. It is worth noting that the system operating point should not exceed *D* in the accelerating phase. In other words, the angle overshoot should not cross *D* that determines maximum buffer area. Otherwise, the operating point would enter adjacent decelerating area, leading to reverse regulation. In such cases, the operating point cannot converge to *C* and LOS would occur.

In particular, if $a_k = 0$ or it is insignificant compared with $U_g$, the operating point will not reach *D* because of the damping effect of PLL's proportional regulator, regardless of where initial state locates. Under this condition, the synchronization is always guaranteed. The above analysis suggests that offset term in $u_{kq}$ makes an unfavorable effect on VSCs' synchronization and it should be responsible for potential LOS events. Besides, it is undeniable that inherent properties of second-order PLL itself cause the phenomenon of angle overshoot, which is another crucial factor of causing LOS. The combined effect of offset term and angle overshoot is the root cause of LOS.

### B. Impact of Multi-Converter Interaction on Stability

Equation (6) yields the offset term $a_k$ as follows,

$$a_k = \mathrm{Im}\left[\underbrace{Z_g \sum_{j=1, j\neq k}^{N} \dot{I}_j^k + \sum_{i=1}^{k} Z_i \sum_{j=i, j\neq k}^{n} \dot{I}_j^k}_{Mutual\ effect} + \underbrace{\left(Z_g + \sum_{i=1}^{k} Z_i + Z_{kt}\right) \dot{I}_k^k}_{Self\ effect}\right] \quad (12)$$

where the mutual effect term quantitatively represents the interaction of $VSC_k$ with the other VSCs. It includes the interactions on grid impedance ($Z_g$) and the interaction on collector line impedance ($Z_i$). These two interactions can be analyzed separately by splitting the mutual effect term, i.e., setting $Z_i = 0$ and $Z_g = 0$ separately. If setting $Z_i = 0$ and taking $n = N = 2$ as an example, the interaction on grid impedance is illustrated in Fig. 9(a). The offset terms annotated in Fig. 9(a) are

$$\begin{aligned} a_1 &= \mathrm{Im}\left[\left(Z_g + Z_{1t}\right)\dot{I}_1^1 + Z_g \dot{I}_2^1\right] \\ a_1' &= \mathrm{Im}\left[\left(Z_g + Z_{1t}\right)\dot{I}_1^1\right] \end{aligned} \quad (13)$$

which indicates that the output current from $VSC_2$ makes an effect through $Z_g$ on $VSC_1$'s input. The effect becomes significant with the increase of grid impedance, which suggests that multi-converter interaction is especially remarkable in high-impedance weak grid. If setting $Z_g = 0$, the interaction on collector line is illustrated in Fig. 9(b),

$$\begin{aligned} a_2 &= \mathrm{Im}\left[\left(Z_1 + Z_2 + Z_{2t}\right)\dot{I}_2^2 + Z_1 \dot{I}_1^2\right] \\ a_2' &= \mathrm{Im}\left[\left(Z_1 + Z_2 + Z_{2t}\right)\dot{I}_2^2\right] \end{aligned} \quad (14)$$

which suggests that the output current from $VSC_1$ affects $VSC_2$'s input through collector line impedance $Z_1$. Note that if collector line interconnects with more VSCs, the interaction behavior will become more complicated. Equations (13) and (14) imply that the interaction on grid impedance and that on collector line impedance have similar effect on the input of VSCs. Here, the former is taken as an example to understand the impact of multi-converter interaction.

*1) Impact on Equilibrium Point*: Since the interaction on grid impedance in (13) enhances the offset term (absolute value), it is easier for two-converter systems than single-converter systems to lose SEP, as shown in Fig. 10. Essen-tially, more VSCs result in larger output current, larger voltage drop across grid impedance and therefore easier to lose SEP.

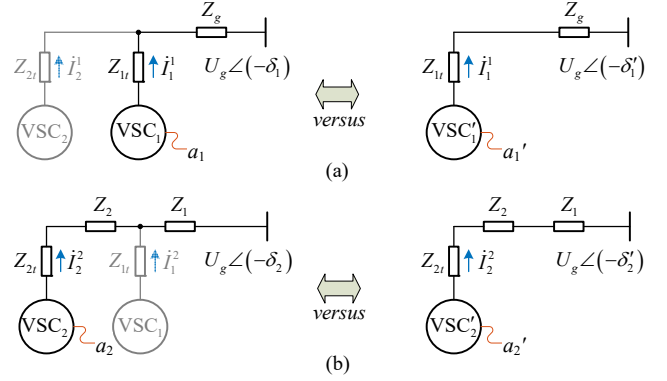

Fig. 9. Multi-converter interaction illustrations: (a) Interaction on grid impedance. (b) Interaction on collector line impedance.

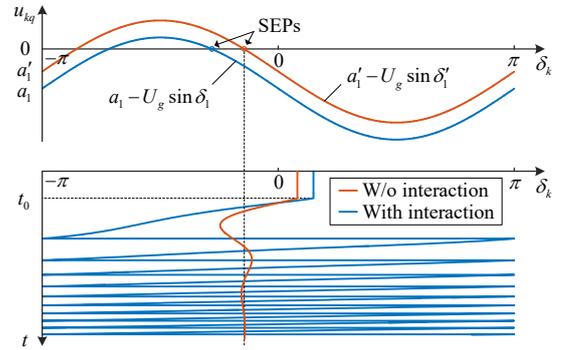

Fig. 10. The two-converter interaction term in (13) makes the offset term (absolute value) larger and hence deteriorates the synchronization stability.

*2) Impact on Dynamic Performance*: Fig. 8 has suggested that the offset term in $u_{kq}$ makes the curve of $u_{kq}$ asymmetric with respect to the axis of $\delta_k$, and consequently there may be not enough buffer area accommodating overshoot of PLL. Fig. 10 shows that the allowable buffer area for overshoot is reduced with the increase of offset term (absolute value). Consequently, the synchronization stability is destroyed and LOS occurs.

## IV. METHODS TO AVOID LOS

### A. Existing Four Methods

*1) PLL Freezing Method* [16]: Fig. 11 displays four existing methods addressing LOS. One of the simplest methods is PLL freezing method, which freezes PLL's proportional-integral (PI) regulator when a grid fault is detected. Once PLL is frozen, it becomes an open-loop system and then autonomously outputs angle according to the recorded frequency and angle at the instant of freezing. Since the frozen PLL can no longer track voltage angle, there is static error at steady state. It should be noted that phase jumps usually appear in grid voltage at the moment of grid faults. The phase jump cannot be detected by frozen PLL. Consequently, current vector control with frozen PLL cannot meet mandatory grid code specifications [10].

*2) Variable Structure PLL (VSPLL) Method* [10]: This method just freezes/removes PLL's integral regulator while reserving proportional regulator, as shown in Fig. 11(b). Thus, it remains regulation capability while avoiding unfavorable angle overshoot caused by second-order PLL. The VSPLL could be seen as a compromise between original second-order PLL and frozen PLL. The method is very simple, but it cannot converge if there is no equilibrium point after grid faults [10].

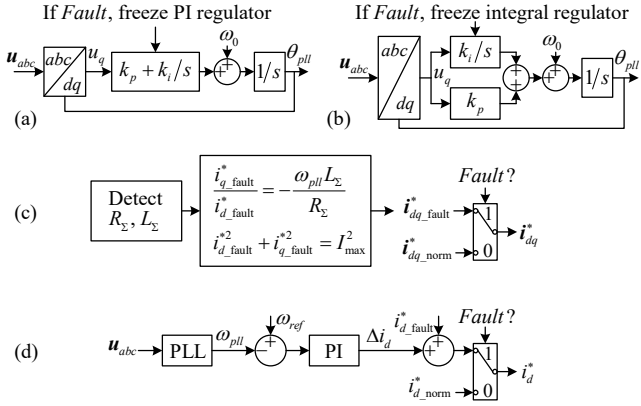

Fig. 11. Four existing typical methods addressing LOS. (a) PLL freezing method [16]. (b) Variable structure PLL method [10]. (c) Adaptive current injecting (ACI) method [8]. (d) PLL frequency based active current regulating method [17].

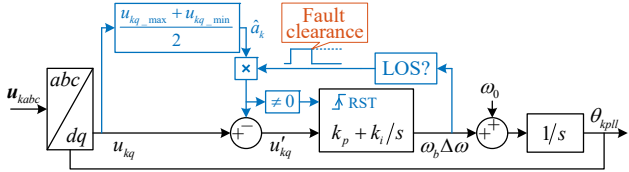

Fig. 12. Proposed feedforward compensated PLL method.

*3) Adaptive Current Injecting (ACI) Method* [8]: This method specifies the ratio of active/reactive current according to the angle of post-fault grid impedance, as shown in Fig. 11(c). To this end, the method requires fast impedance angle estimation. Besides, the method is developed from the standpoint of SCIB system. The current references in Fig. 11(b) yield $a = 0$ in (4), and thus it eliminates the effect of grid impedance interaction. However, the method becomes ineffective for multi-converter systems because the composition of $a_k$ [see (7)] is much complex and it cannot be eliminated using the method.

*4) PLL Frequency Based Active Current Regulating Method* [17]: This method in Fig. 11(d) regulates active current to avoid LOS. Reference [17] verified the effectiveness of the method by simulations instead of in theory. The steady state of the method seems to be unpredictable. For the simplest SCIB system, $u_q = 0$ at steady state yields that

$$R_\Sigma i^*_{q\_fault} + \omega_{pll} L_\Sigma \left(i^*_{d\_fault} + \Delta i_d\right) - U_g \sin \delta = 0 \quad (15)$$

where $\Delta i_d$ and $\delta$ are two variables, but there is only one equation. All possible steady state equilibrium points form a curve. It is difficult to predict the point where the system stabilizes ultimately because of being affected by initial value, control parameters, etc. Moreover, it seems to be difficult to theoretically prove the stability for multi-converter systems with the method.

To sum up, these four methods have various disadvantages affecting their performance. Also, their applicability for multi-converter systems is in doubt or invalid. To this end, a novel method applicable for multi-converter systems is developed.

*B. Proposed Feedforward Compensated PLL*

Recalling the root cause of LOS, it involves two factors: offset term in $u_{kq}$ and PLL's overshoot property. Eliminating either of the twofold factors is helpful to address the issue of LOS. This study proposes to use a feedforward compensation to eliminate the effect of the first factor. The method is called feedforward compensated PLL (FFC-PLL). Briefly, if the offset term in $u_{kq}$ is estimated by some means, its effect can be eliminated by subtracting the estimation value from $u_{kq}$.

When LOS occurs, Equation (7) indicates that $u_{kq}$ shows a continuous oscillation. The maximum and minimum of the oscillation are

$$\begin{aligned} u_{kq\_max} &= a_k + U_g \\ u_{kq\_min} &= a_k - U_g \end{aligned} \quad (16)$$

Since $u_{kq}$ as input signal of PLL is detectable and therefore it is knowable, the offset term $a_k$ could be estimated by,

$$\hat{a}_k = \frac{u_{kq\_max} + u_{kq\_min}}{2} \quad (17)$$

Adding $-\hat{a}_k$ as a feedforward compensation into $u_{kq}$ (see Fig. 12) can eliminate the effect of the offset term. Consequently, it makes an ideal form of PLL input with zero offset:

$$u'_{kq} = 0 - U_g \sin \delta_k \quad (18)$$

which can ensure the synchronization of VSCs.

Fig. 12 shows the feedforward compensated PLL. Potential LOS is identified by means of a dead zone of frequency deviation after VSCs enter fault ride-through stage. The feedforward compensation is enabled only if frequency deviation is large enough, which suggests that LOS really occurs. If synchronization has been achieved rapidly after a slight grid fault, there is no risk of LOS and thus it is unnecessary to activate the feedforward compensation. Furthermore, a resetting operation at rising edge is applied to PI regulator in order to remove the frequency deviation and quicken synchronization. After fault clearance, the feedforward compensation would be deactivated.

After LOS occurs, the estimation by (17) needs at least half a cycle of oscillation time to grasp the maximum and minimum of $u_{kq}$. From (7) and (8), it can be obtained that

$$u_{kq} = a_k - U_g \sin\left[\int \omega_b \left(\omega_{kpll} - \omega_g\right) dt + \delta_{k0}\right] \quad (19)$$

which indicates that the faster PLL frequency deviates from grid frequency, the greater oscillation frequency ($\omega_{kpll} - \omega_g$), and consequently the faster estimation of the offset term.

V. SIMULATION RESULTS

Simulations are conducted on MATLAB/Simulink for verifying the performance of the proposed FFC-PLL. In order to simulate actual converter as far as possible, the model of converter in Fig. 1 is built with switch-level electromagnetic transients. The parameters of the model are summarized in Table I.

Two cases are set in simulation verifications, including Case A and Case B. Firstly, the performance of the proposed FFC-PLL is verified with a single-VSC system in Case A. Then, the effectiveness of the method for a multi-VSC system is verified in Case B, in which a two-VSC system (see Fig. 6) is taken as an example. Note that the two-VSC system could also be a representation of two VSC interfaced generation plants. With this consideration, line lengths of $Z_1$ and $Z_2$ are set to 5 km and 50 km, respectively. The line impedance $Z_2$ makes the operating points of two VSCs distinguished.

The offset term in $q$-axis voltage can be calculated with the model parameters in Table I. Table II shows the theoretical true value. Figs. 11 and 12 show simulation results. After LOS is detected in simulations, the offset term can be estimated by (17). Table II also shows the estimated value. The relative error of the estimation is given in the last column, which suggests that the estimation has high enough accuracy.

### A. Verification on a Single-VSC System

Fig. 13 displays simulation result of a single-VSC system. The VSC outputs $1.0 + 0.0j$ p.u. current in normal operating condition. Grid voltage dips to 0.05 p.u. at 0.2 s, which causes the terminal voltage $U$ to sag, as shown in Fig. 13(b). The VSC outputs $0.0 - 1.0$ p.u. current during grid faults from 0.2 to 0.7 s. Since the offset term (absolute value ~0.1 p.u.) during this period is larger than the residual grid voltage (0.05 p.u.), there is no post-fault equilibrium point. As a consequence, LOS occurs, as shown in Fig. 13(e), in which PLL frequency deviates from the fundamental frequency. Fig. 13(d) shows that $u_q$ exhibits a continuous oscillation while utilizing original PLL. After experiencing maximum and minimum points, the offset term can be estimated.

According to the schematic diagram in Fig. 12, the opposite of the estimated offset term is added into $u_q$, as shown in Fig. 13(d). Meanwhile, a resetting operation is applied to PI regulator of PLL. With this compensation added, the adverse effect of offset term is eliminated. Therefore, the VSC can rapidly achieve synchronization, as shown in Fig. 13(e). The performance comparison between the proposed FFC-PLL and original PLL verify the feasibility of FFC-PLL addressing LOS.

### B. Verification on a Multi-VSC System

Fig. 14 displays simulation result of a two-VSC system. The total capacity of the system is twice as large as the single-VSC system capacity. Hence, both VSCs outputs $0.5 + 0.0j$ p.u. current in normal operating condition and $0.0 - 0.5$ p.u. current during grid faults. The grid fault is the same as that in Case A.

TABLE I
PARAMETERS OF SIMULATION MODEL

| Capacity | 2 MW | Transformer | 0.002+0.05j p.u. |
|---|---|---|---|
| Rated voltage | 690 V | Line resistance | 0.1153 Ω/km |
| Frequency | 50 Hz | Line inductance | 0.3299 Ω/km |
| Fault voltage | 0.05 p.u. | PLL | 150+2500/s |
| Grid impedance | 0.1+0.3j p.u. | Current loop | 2+10/s |
| Output filter | 0.15 p.u. | Sample time | 5 μs |

TABLE II
RELATIVE ERROR OF THE ESTIMATED OFFSET TERM BY (17)

| | True value | Estimated value | Relative error |
|---|---|---|---|
| Case A VSC | 0.1029 p.u. | 0.1001 p.u. | 2.8% |
| Case B VSC1 | 0.1039 p.u. | 0.1043 p.u. | 0.4% |
| Case B VSC2 | 0.1133 p.u. | 0.1070 p.u. | 5.6% |

Fig. 14(e) shows that LOS happens after the grid voltage sag. After feedforward compensations are estimated and added into $u_{1q}$ and $u_{2q}$, both PLLs can synchronize with the grid, as shown in Fig. 14(e). Since operating points of two VSCs are different, the estimated offset terms are accordingly different. Moreover, according to the analysis in Section III.B, the offset term estimated by (17) contains the self-effect and mutual effect (i.e., multi-converter interaction). Therefore, the cumulative effect from both effects can be eliminated once the feedforward compensation is performed, which fundamentally explains why this method is suitable for multi-converter systems.

### C. Comparisons with Existing Methods

Fig. 15 shows comparisons with four existing methods. Three methods including FFC-PLL, adaptive current injecting, and PLL frequency based active current regulating, all can achieve synchronization. Strictly, however, the latter two methods have the problem of effectiveness for multi-converter systems (as remarked in Section IV.A). It is found in Fig. 15(b) that PLL freezing method has a fatal problem with static error ($u_q \neq 0$). In other words, it actually outputs a wrong angle. It is also observed that variable structure PLL fails to converge, which is because there is no equilibrium point. The above comparisons indicate the advantage of the proposed FFC-PLL.

## VI. CONCLUSIONS

It is much significant but not easy for VSCs connected to a high-impedance weak grid to synchronize with the grid during severe grid faults. Potential loss of synchronism (LOS) probably leads plentiful VSCs in a wide area to disconnect from the power grid, which violates grid codes and endangers the stability of the power grid. The synchronization stability of multi-converter systems was very little studied previously. For this, this paper is devoted to filling this gap. The contribution includes two points: i) The model of multi-converter infinite-bus system was developed. The cause of LOS in multi-converter systems was clarified. The existing knowledge of single-converter systems in this respect was extended to multi-converter systems. ii) A feedforward compensated PLL method applicable for multi-converter systems was proposed to address LOS.

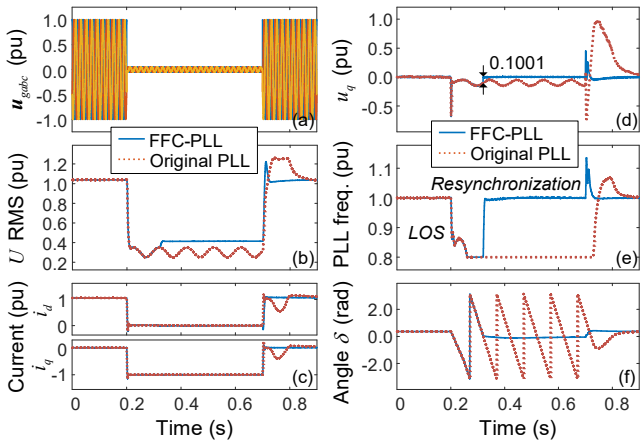

Fig. 13. Simulation result of a single-converter system.

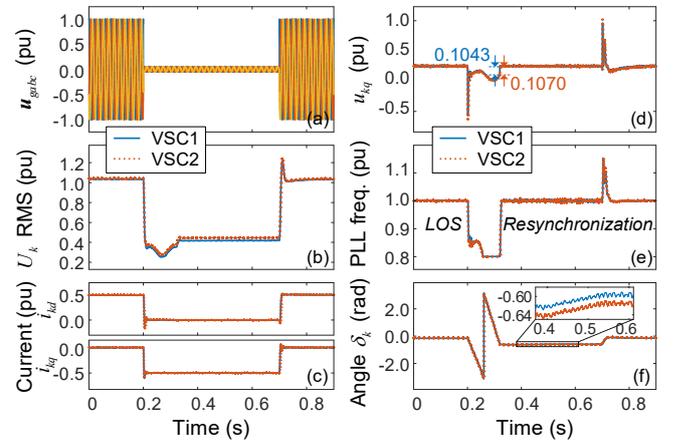

Fig. 14. Simulation result of a two-converter system.

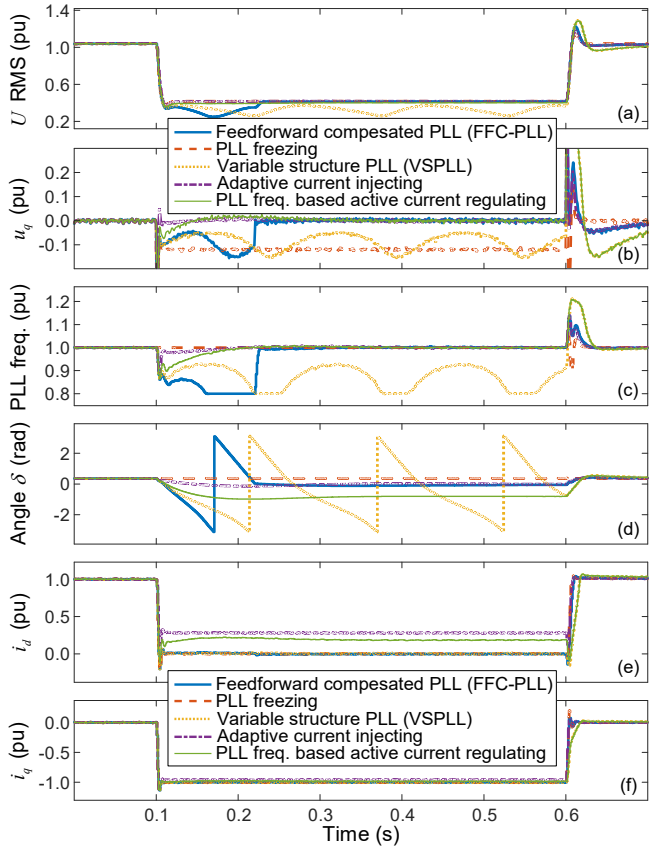

Fig. 15. Comparisons with existing four methods.

It has been found that LOS in multi-converter systems is because of two crucial factors. One is voltage drop across grid impedance. It causes an offset term in PLL input and potentially leads to absence of equilibrium point after grid faults. The other is PLL's inherent overshoot property. The proposed feedforward compensated PLL is dedicated to eliminating the effect of the first factor. The performance of the method has been analyzed and compared with four existing methods. Because of simplicity and applicability for multi-converter systems, this method has good potential to address the issue of LOS in industrial practice.

ACKNOWLEDGMENTS

This work is supported by the National Natural Science Foundation of China (Nos. 61722307, 5191101838).